\documentclass[pdflatex,sn-nature]{sn-jnl}

\usepackage{graphicx}%
\usepackage[strings]{underscore}
\usepackage{multirow}%
\usepackage{amsmath,amssymb,amsfonts}%
\usepackage{amsthm}%
\usepackage[title]{appendix}%
\usepackage{xcolor}%
\usepackage{textcomp}%
\usepackage{manyfoot}%
\usepackage{booktabs}%
\usepackage{algorithm}%
\usepackage{algorithmicx}%
\usepackage{algpseudocode}%
\usepackage{listings}%
\usepackage{lmodern}
\usepackage{url}
\usepackage{lineno}

\begin{document}

\title[Article Title]{Anticipating AMOC transitions via deep learning}

\author[1,2]{\fnm{Wenjie} \sur{Zhang}}

\author*[2,3]{\fnm{Yu} \sur{Huang}}\email{y.huang@tum.de}

\author[2,3]{\fnm{Sebastian} \sur{Bathiany}}

\author[4]{\fnm{Yechul} \sur{Shin}}

\author[2,3]{\fnm{Maya} \sur{Ben-Yami}}

\author[1]{\fnm{Suiping} \sur{Zhou}}

\author[2,3,5]{\fnm{Niklas} \sur{Boers}}

\affil[1]{\orgdiv{School of Aerospace Science and Technology}, \orgname{Xidian University}, \city{Xi’an}, \country{China}}

\affil[2]{\orgdiv{Earth System Modelling}, \orgdiv{School of Engineering and Design}, \orgname{Technical University of Munich}, \city{Munich}, \country{Germany}}

\affil[3]{\orgdiv{Complexity Science}, \orgname{Potsdam Institute
for Climate Impact Research}, \city{Potsdam}, \country{Germany}}

\affil[4]{\orgdiv{School of Earth and Environmental Sciences}, \orgname{Seoul National University}, \city{Seoul}, \country{Republic of Korea}}

\affil[5]{\orgdiv{Department of Mathematics and Global Systems Institute}, \orgname{University of Exeter}, \city{Exeter}, \country{UK}}

\abstract{Key components of the Earth system can undergo abrupt and potentially irreversible transitions when the magnitude or rate of external forcing exceeds critical thresholds. 
In this study, we use the example of the Atlantic Meridional Overturning Circulation (AMOC) to demonstrate the challenges associated with anticipating such transitions when the system is susceptible to bifurcation-induced, rate-induced, and noise-induced tipping. Using a calibrated AMOC box model, we conduct large ensemble simulations and show that transition behavior is inherently probabilistic: under identical freshwater forcing scenarios, some ensemble members exhibit transitions while others do not. In this stochastic regime, traditional early warning indicators based on critical slowing down are unreliable in predicting impending transitions. To address this limitation, we develop a convolutional neural network (CNN)-based approach that identifies higher-order statistical differences between transitioning and non-transitioning trajectories within the ensemble realizations. This method enables the real-time prediction of transition probabilities for individual trajectories prior to the onset of tipping. Our results show that the CNN-based indicator provides effective early warnings in a system where transitions can be induced by bifurcations, critical forcing rates, and noise. These findings underscore the potential in identifying safe operating spaces and early warning indicators for abrupt transitions of Earth system components under uncertainty.}

\maketitle
\section{Introduction}\label{sec1}
Earth System Models (ESMs) and paleoclimate observations suggest that some major components of the Earth system are susceptible to abrupt transitions \cite{dakos2008slowing, armstrong2022exceeding, boers2022theoretical, flores2024critical}, sometimes labeled as 'tipping' behavior. Major candidates of 'tipping elements' in the Earth system are the Amazon rainforest \cite{boulton2022pronounced}, the continental ice sheets \cite{garbe2020hysteresis}, and the Atlantic Meridional Overturning Circulation (AMOC) \cite{boers2021observation, van2024physics}. Previous studies have identified three primary mechanisms that can drive such transitions in nonlinear dynamical systems \cite{ashwin2012tipping, brunetti2023attractors, chapman2024tipping}. First, a gradual increase in external forcing may exceed a critical threshold, triggering a bifurcation-induced transition and thus causing a sudden shift to an alternative equilibrium \cite{dakos2008slowing, scheffer2009early, Kleinen2003, held2004detection}. Second, if the rate of external forcing is sufficiently rapid compared to the intrinsic time scale of the system, the system may not remain within the basin of attraction of its original equilibrium, leading to a loss of stability and a subsequent transition \cite{lohmann2021risk, alharthi2023rate, huang2024deep}. Third, persistent stochastic perturbations can drive the system across basin boundaries even in the absence of changes in external forcing, leading to noise-induced transitions between alternative equilibria \cite{hasselmann1976stochastic, ashwin2012tipping, timmermann2000noise, stocker1997influence,boers2018early, cini2024simulating}. The rapid pace of anthropogenic forcing compared with the internal timescales of key Earth system components, such as the AMOC, poses significant risk and uncertainty associated with abrupt transitions from combined effects of these three mechanisms \cite{lenton2019climate, ritchie2021overshooting, wunderling2023global, romanou2023stochastic, klose2024rate, sinet2024amoc, baker2025continued}. Therefore, identifying the boundaries of safe operating spaces and early warning signals of potential state transitions are important and challenging tasks for the study of future climate change and climate mitigation policy \cite{lenton2019climate, ritchie2021overshooting, armstrong2022exceeding, ben2024uncertainties, rietkerk2025ambiguity}. Safe operating spaces define the boundaries within which a system can function without undergoing undesired transitions.

The dominant approach to understanding the mechanisms and characteristics of critical transitions of the AMOC involves the use of process-based models. These models investigate AMOC instability and critical thresholds in the context of external forcing and internal climate variability. A common experimental setup applies a gradually changing external forcing to the AMOC until a critical threshold is reached, beyond which bifurcation-induced transition typically occurs \cite{zhang2017abrupt, romanou2023stochastic, van2024physics, ben2023uncertainties, ditlevsen2023warning, dijkstra2024role}. However, recent ensemble simulation studies have demonstrated that under identical scenarios of increasing anthropogenic forcing, different internal climate variability between realizations can lead to divergent outcomes \cite{romanou2023stochastic, lohmann2024multistability, cini2024simulating, gu2024wide, curtis2024collapse}. In some realizations, the AMOC undergoes sustained weakening followed by tipping, while in others, transitions do not occur. Because bifurcation-induced transitions are inherently deterministic, these results suggest that rate- and noise-induced effects can introduce stochasticity into the onset of AMOC tipping. A follow-up question is whether this mechanism also introduces uncertainty to the critical thresholds and safe operating spaces. Despite its importance, conducting large ensemble simulations in comprehensive Earth system models to systematically investigate the stochastic nature of AMOC transitions is currently difficult to achieve due to the high computational cost. As an alternative, conceptual box models offer a more efficient framework for capturing the essential dynamics governing AMOC stability \cite{frierson2013contribution, wood2019observable, dijkstra2024role, chapman2024tipping}. Recent advances have utilized simulation data from comprehensive ESMs to calibrate a five-box model of the AMOC \cite{wood2019observable, chapman2024tipping}, with emphasis on the dynamical coupling and feedback mechanisms among the oceanic components of Northern Atlantic, Tropical Atlantic, North Atlantic Deep Water, Indo-Pacific, and Southern Ocean. The calibration enables the models to reproduce the double equilibria and critical transitions characteristic of the AMOC in comprehensive ESMs \cite{wood2019observable, alkhayuon2019basin, dijkstra2024role, chapman2024quantifying, chapman2024tipping, panahi2024machine, sinet2024amoc, pals2024targeted}. In this study, we employ this calibrated box model of Wood et al. \cite{wood2019observable} to generate a large ensemble of simulations that we use to investigate the critical thresholds  when the system is susceptible to bifurcation-induced, rate-induced, and noise-induced tipping. 

The collective behavior of a large ensemble of simulated realizations provides insights into the characteristics and uncertainties associated with tipping events. In practical applications, however, it is often necessary to identify early warning signals of pending transitions from a single noisy time series \cite{scheffer2009early, lenton2011early, ben2024uncertainties, rietkerk2025ambiguity}. A key challenge, therefore, is to extract predictive insights from the empirical distribution of ensemble trajectories to inform the prediction of tipping in individual time series. Critical slowing down (CSD) theory suggests that as the local stability of a system weakens, even minor perturbations can have a prolonged effect on the system's dynamics \cite{dakos2008slowing, scheffer2009early, lenton2011early, rietkerk2025ambiguity}. This is typically reflected in increased variance and lag-1 autocorrelation of the noisy time series, serving as statistical early warning indicators preceding bifurcation-induced tipping. Nonetheless, CSD-based indicators are ineffective in detecting rate- and noise-induced tipping \cite{huang2024deep}, whereas deep learning approaches \cite{bury2021deep, bury2023predicting, huang2024deep, panahi2024machine} have shown enhanced predictive performance in predicting rate-induced and noise-induced tipping \cite{huang2024deep}. However, the neural networks used in Huang et al. 2024 \cite{huang2024deep} have been constrained to predict transitions only with fixed forecast lead times. Moreover, the applicability of deep learning predictions has not yet been validated under scenarios involving combinations of the three types of tipping mechanisms. Here we  aim to improve upon these deep learning approaches by removing the reliance on predefined lead times, thereby enabling real-time prediction of AMOC tipping probability. We demonstrate that convolutional neural networks (CNNs) can extract features related to higher-order statistical moments from the empirical distribution of ensemble realizations, allowing for the prediction of tipping events in a single time series under the combined effects of the three types of tipping mechanism.

\section{Results}\label{sec2}
\subsection{Stochastic abrupt transition under critical magnitude and rate of forcing}\label{subsec2}
We begin by analyzing the deterministic dynamics of AMOC instability in the absence of stochastic perturbations, using a box model calibrated to the comprehenisve ESM HadGem3-GC31-MM \cite{chapman2024quantifying, chapman2024tipping}. By gradually increasing and then decreasing the freshwater forcing, we identify the bifurcation diagram at which the AMOC undergoes abrupt transitions between distinct equilibrium states (Supplementary Fig. S1). Supplementary Fig. S2a illustrates bifurcation- and rate-induced tipping, respectively, under noise-free conditions. Specifically, under a slowly increasing forcing from 0 to 0.6 Sv over 10,000 years, the AMOC abruptly tips once the forcing exceeds 0.53 Sv, indicating the critical threshold of bifurcation-induced tipping in case of the deterministic dynamics (Supplementary Fig. S2b). In contrast, a rapidly changing forcing can also trigger the transition at a lower forcing value \cite{lohmann2021risk, chapman2024tipping}. For instance, when the freshwater forcing rapidly increases from 0 to 0.41 Sv over just 200 years (Supplementary Fig. S2c), the system also collapses, which is a case of rate-induced tipping. This emphasizes that both the magnitude and rate of forcing determine when AMOC instability occurs \cite{lohmann2021risk}. To systematically identify the critical thresholds for both bifurcation- and rate-induced effects, we perform simulations across a broad range of forcing scenarios, with different combinations of the forcing magnitude and its changing rate. As a result, the deterministic boundary distinguishing tipping from non-tipping conditions is defined by joint thresholds from the bifurcation- and rate-induced effects (Supplementary Fig. S2d).   

To incorporate the stochastic nature, the box model is further extended by subjecting it to noise perturbations derived from the HadGem3-GC31-MM \cite{chapman2024quantifying, chapman2024tipping} (see Methods). To account for uncertainty arising from stochastic dynamics, we conduct 20,000 ensemble simulations, each subjected to identical freshwater forcing but with independent realizations of white noise. In the slow forcing scenario, where freshwater forcing increases from 0 to 0.6 Sv over 10,000 years, all realizations undergo tipping when the forcing exceeds an average threshold of 0.43 Sv (Figs. 1a and 1c). However, under a fast forcing scenario (increasing from 0 to 0.39 Sv over 200 years), only 21\% of the realizations tip while the others remain stable (Figs. 1b and 1d) - thus the tipping occurrence is stochastic. Fig. 1e presents the proportion of the tipping realizations within ensemble simulations at each combination of the forcing magnitude and its rate of change. This reveals a nonlinear and probabilistic transition boundary that depends not only on the magnitude of the forcing but also on its rate, suggesting that the boundary of the AMOC's safe operating space is not a fixed threshold, but rather a probabilistic zone. Furthermore, this result demonstrates that even forcing below deterministic tipping thresholds can lead to tipping with substantial probability under the combined effects of bifurcation-, rate- and noise-induced mechanisms. The contrasting critical threshold patterns between deterministic and stochastic dynamics are shown in Fig. 1e and Supplementary Fig. 2d. Consequently, the boundaries of the system's “safe operating space” become smeared out. 

\subsection{Large uncertainty for canonical early warnings}
We examine the CSD indicators and AMOC strength by testing them on the ensemble realizations under a representative freshwater forcing scenario. This scenario is characterized by a linear increase in forcing with an upper magnitude of 0.42 Sv over 800 years (Fig. 2a). Under this forcing scenario, tipping can be triggered through a combination of bifurcation-, rate-, and noise-induced tipping mechanisms (Fig. 1e). From the ensemble simulation, we extract 50,000 time series that exhibit tipping, where tipping is defined as the AMOC strength reaching 0 Sv, and the tipping time as the moment this threshold is first reached. All realizations that pass the 0 Sv eventually reach the weak AMOC state. In addition, we also pick 50,000 time series that do not tip, creating a balanced dataset of 100,000 samples. The ensemble time series in the tipping category exhibit a wide range of tipping times (Fig. 2a). This is consistent with recent findings that considerable uncertainty persists in estimating tipping times \cite{ben2024uncertainties, curtis2024collapse}. For subsequent analysis, each time series is detrended by subtracting the ensemble mean of the non-tipping time series, then we align the ends of all time series to their tipping times (Methods). As expected from theory, the computed CSD indicators and the detrended AMOC strength as an indicator, show substantial overlap between the tipping and non-tipping categories (Supplementary Figs. S3a-S3c), indicating they cannot be used to anticipate the transitions. Indeed, the comparison of receiver operating characteristic (ROC) score \cite{FAWCETT2006861} suggests that all of these canonical early warning indicators have no prediction skill at forecast lead times of 200 years before the tipping point (Figs. 2b, 2c, and Supplementary Fig. S4a). Although the AMOC strength typically decreases in the tipping realizations, the differences between the tipping and non-tipping ensembles remain statistically indistinguishable until shortly before the tipping point. 

\subsection{Disentangling the predictive signal with ensemble distributions}
We further examine the ensemble probability distributions using a lead time of 200 years as an illustrative example. Although the probability density functions (PDFs) of the tipping and non-tipping categories overlap in their spreads, their distributional profiles reveal subtle differences in their kurtosis (Fig. 3a), suggesting that distinctions between the two groups of data may persist in higher-order statistical moments or within higher-dimensional feature spaces \cite{scholkopf1997kernel, he2013detecting, huang2024deep}. 

We develop a CNN-based classifier (Methods) to discriminate potential differences between time series with and without transitions. In previous studies, deep learning models were typically trained at a specific lead time prior to the onset of tipping events. Such approaches are inherently limited by their reliance on fixed forecast lead times, which constrains their applicability across varying temporal scenarios \cite{huang2024deep}. As illustrated in Fig. 4a, to address this limitation, we adopt a sliding-window strategy to extract multiple fixed-length time series segments preceding the tipping point for CNN training. Each segment has a fixed window length of 200 years, and the sampling windows slide backward along the time axis from the tipping point to capture features preceding a potential tipping event. In this study, we fix the sampling range to three partially overlapping windows, each 200 years long, with adjacent windows overlapping by 100 years, as shown in Fig. 4a (windows-3-stride-100). During sampling, all segments extracted from the same realization are assigned the same label: 1 if the realization undergoes a tipping event, and 0 otherwise. These time series segments are treated as independent training samples, allowing the neural networks to learn early-warning patterns across different lead times. Once trained, the neural networks can take any 200-year time series segment as input, and output the estimated probability of an impending transition. Fig. 4b shows that the neural networks can make skillful prediction across different forecast lead times. This approach allows the neural networks to learn from a wider range of precursory dynamics and removes the constraint of requiring inputs to align with a specific forecast lead time during inference. We test a range of windowing strategies that extract time series segments preceding the tipping time with varying lengths and strides (Fig. 4 and Supplementary Fig. S5). Among these, the windows-3-stride-100 (window length=200 years) configuration yields the best predictive performance, particularly for forecast lead times between 100 and 200 years (Fig. 4b). Based on this result, we adopt the windows-3-stride-100 setting for all subsequent experiments unless otherwise specified. 

To visualize the CNN-based classifier's decision making process, we employ layer-wise relevance propagation (LRP), which identifies the attention weights most relevant to the predictions of neural networks in the corresponding target samples \cite{bach2015pixel, montavon2019layer, huang2024deep} (Methods). Fig. 3b presents the joint probability density of detrended AMOC strengths and the LRP attention scores from the trained CNN classifier, where the tipping and non-tipping categories become separable. This demonstrates that the difference between tipping and non-tipping categories can be discriminated with a considerable lead time (see also Fig. 3e).

\subsection{A deep learning-based real-time indicator}\label{subsec4}
 From the above results we infer that the trained neural networks can estimate the probability of an impending AMOC tipping event in real time, without being constrained to a predefined forecast lead time. We apply the trained neural networks to continuously monitor the tipping probability based on the evolving AMOC strength time series. As shown in Fig. 2d, the average predicted tipping probability for tipping and non-tipping ensembles begins to diverge as early as year 700 in a 2800-year time series. For tipping realizations, the average predicted probability steadily increases and approaches 1 around year 1500, while the non-tipping ensemble maintains a low and nearly constant probability below 0.2. This temporal separation demonstrates that the model is able to track evolving system dynamics and provide real-time early warnings. As illustrated in Supplementary Fig. S3d, approximately 100 years before the actual transition, the predicted tipping probability for the tipping realizations exceeds the 99\% confidence interval of the non-tipping ensemble, providing a very early and statistically robust prediction signal. Accordingly, the ROC analysis (Supplementary Fig. S4) shows that the CNN-based indicator outperforms the traditional AMOC strength–based indicator at forecast lead times of 200 and 150 years, where the CSD indicators exhibit no predictive skill.  

\subsection{Warning of tipping under combined bifurcation, rate and noise effects}\label{subsec5}
To assess the feasibility of predicting stochastic tipping of AMOC in the general setting of combined bifurcation-, rate- and noise-induced effects, we generate a large ensemble of realizations for each freshwater forcing scenario shown in Fig. 1e. Each scenario is parameterized by the magnitude and timescale of freshwater input, spanning a regime where tipping outcomes are intrinsically stochastic. In the following, two sets of experiments focusing on scenarios with stochastic tipping proportions between 5\% and 95\% at a 150-year forecast lead time. 

We first examine whether the foresaid neural networks, trained solely on data from a single freshwater forcing scenario (linear increase to 0.42 Sv over 800 years), could generalize its prediction capability to other forcing scenarios. Despite this limited training domain, our neural network can achieve high accuracy across a wide region of forcing scenarios (Fig. 5a), particularly in scenarios characterized by strong forcing magnitudes and relatively slow forcing rates. These are conditions under which bifurcation- and noise-induced tipping dominate, and our findings suggest that neural networks extract robust precursors invariant to the specifics of forcing. However, performance degrades under fast forcing scenarios, likely due to the shortened pre-tipping trajectories that limit the available predictive information and the increased influence of rate-induced dynamics, which are harder to detect from early-time signals. As a benchmark, we additionally train a second CNN classifier using a balanced dataset sampled across all relevant forcing scenarios. For each scenario with a tipping probability between 5\% and 95\%, we draw 1,000 tipping and 1,000 non-tipping trajectories, aggregating them to form a comprehensive training dataset. The neural networks trained under this broader distribution demonstrated enhanced generalization, as shown in Fig. 5b, achieving higher accuracy even in the rapid-forcing regimes where traditional statistical early warning indicators tend to fail. Notably, prediction accuracy remains robust even near the boundaries of the stochastic tipping regime, where the nonlinear intersection among bifurcation-, rate-, and noise-induced effects introduces substantial complexity and uncertainty. 

\section{Discussion}\label{sec3}
Our study provides insights into the predictability of AMOC transitions in the general setting of combined  bifurcation-, rate-, and noise-induced effects. Leveraging a recent AMOC box model calibrated to emulate a comprehensive ESM \cite{wood2019observable}, we conduct large ensembles of stochastic simulations under diverse freshwater forcing scenarios. These simulations suggest that the safe boundary for AMOC collapse is not defined by deterministic thresholds, but instead emerges as a probabilistic zone due to internal variability, with the risk of transition further shaped by the magnitude and rate of external forcing, as well as stochastic perturbations. Critically, focusing only on the equilibrium case and assocaited CSD-based warning signals may lead to a severe underestimation of the actual tipping risk and associated false estimation of a safe operating space. 

While bifurcation-induced transitions display early warning signals consistent with classical CSD, such indicators prove unreliable under rate- or noise-induced tipping. In contrast, our deep learning framework, which is trained on empirical distributions of ensemble realizations, overcomes this limitation by enabling real-time prediction of the tipping probability without requiring a predefined forecast lead time and without being confined to just one of the possible underlying tipping mechanisms (bifurcation, rate, or noise). Comparative analysis of detrended AMOC strength, LRP attention maps, and CNN-predicted tipping probabilities confirms that the neural networks capture emergent features beyond the low-order statistical moments used for detecting CSD and other classical early warning signals. 

Our findings have two key implications. First, we show that the concept of a deterministic critical threshold may be inadequate for informing tipping risk assessments targeting the AMOC. Instead, probabilistic boundaries of safe operating spaces, derived from ensemble-based statistics, offer a more comprehensive framework for evaluating system stability. Second, our approach provides a scalable method for developing early warning indicators that account for the combined effects of bifurcation-, rate-, and noise-induced tipping. This methodology has broader applicability beyond the AMOC, extending to other Earth system components characterized by multistability and noise sensitivity, such as the Greenland Ice Sheet and the Amazon rainforest \cite{lohmann2024multistability, bochow2023overshooting, flores2024critical}. 

There remains scope for additional future work. While the box model captures essential AMOC dynamics, is calibrated with respect to a comprehensive ESM, and facilitates large-scale simulation, it necessarily abstracts spatial processes and feedbacks represented in full-complexity models \cite{shin2024fast}. Further integration with intermediate-complexity or fully coupled climate models may help validate the robustness of our findings and assess their sensitivity to model architecture. Moreover, while LRP offers important interpretability, further integration with causal discovery tools \cite{spirtes2016causal} or physics-informed neural networks \cite{RAISSI2019686} might improve mechanistic attribution.

In summary, our results highlight the need to refine conceptual and computational tools for anticipating tipping events in the climate system. As anthropogenic forcing continues to accelerate, integrated approaches that blend dynamical theory, reduced-complexity modeling, and deep learning will be essential for managing deep uncertainty at the edge of climatic instability.

\section{Methods}\label{sec4}
\subsection{AMOC box model}\label{subsec2}
To efficiently explore AMOC tipping dynamics under stochastic forcing, we adopt the reduced three-box model of the Atlantic Meridional Overturning Circulation (AMOC) introduced by Chapman et al. \cite{chapman2024tipping}, which is a modification of the five-box model proposed by Wood et al. \cite{wood2019observable}. The reduced three-box model retains the essential dynamics of the full five-box model, including bifurcation-, noise-, and rate-induced tipping behavior. It still consists of five conceptual oceanic boxes of the Northern Atlantic, Tropical Atlantic, Southern Ocean, North Atlantic Deep Water and Indo-Pacific. Following Chapman et al. \cite{chapman2024tipping}, the five-box model is reduced by assuming constant salinity in the Southern Ocean and North Atlantic Deep Water boxes, with values inferred from calibration to HadGem3-GC31-MM. A global salt conservation constraint is then applied to analytically determine the salinity in the Indo-Pacific box. This yields a two-dimensional system governing the evolution of salinity in the Northern Atlantic and Tropical Atlantic boxes. The AMOC strength $q$ is modeled as: 
\begin{equation}
    q = \lambda(\alpha(T_S-T_{Nor.})+\beta(S_{Nor.}-S_S))
\end{equation}
where $\alpha$ and $\beta$ are linear coefficients from a simplified equation of state for seawater, and $\lambda$ is a model parameter determined by the calibration. $T$ and $S$ represent box-averaged temperature and salinity, respectively. The subscripts indicate the box labels, where $S$ denotes the Southern Ocean box and $Nor.$ denotes the Northern Atlantic box. Temperatures are held fixed in the simulations, so salinity dynamics are governed by salinity evolution. The salinity evolution equations differ depending on the sign of $q$.

For $q\geq0$:
\begin{equation}
    V_{Nor.}\frac{dS_{Nor.}}{dt}=q(S_{Trop.}-S_{Nor.})+K_{Nor.}(S_{Trop.}-S_{Nor.})-F_{Nor.}S_{0} 
\end{equation}
\begin{equation}
    V_{Trop.}\frac{dS_{Trop.}}{dt}=q(\gamma S_{S}+(1-\gamma)S_{IP}-S_{Trop.})+K_{S}(S_{S}-S_{Trop.})+K_{Nor.}(S_{Nor.}-S_{Trop.})-F_{Trop.}S_{0} 
\end{equation}

and for $q<0$:
\begin{equation}
    V_{Nor.}\frac{dS_{Nor.}}{dt}=\left| q \right|(S_{B}-S_{Nor.})+K_{Nor.}(S_{Trop.}-S_{Nor.})-F_{Nor.}S_{0} 
\end{equation}
\begin{equation}
    V_{Trop.}\frac{dS_{Trop.}}{dt}=\left| q \right|(S_{Nor.}-S_{Trop.})+K_{S}(S_{S}-S_{Trop.})+K_{Nor.}(S_{Nor.}-S_{Trop.})-F_{Trop.}S_{0} 
\end{equation}
Salt conservation ensures the total salinity content $C$ remains constant:
\begin{equation}
    C=V_{Nor.}S_{Nor.}+V_{Trop.}S_{Trop.}+V_{S}S_{S}+V_{IP}S_{IP}+V_{B}S_{B}
\end{equation}
allowing the salinity of Indo-Pacific box to be written as:
\begin{equation}
    S_{IP}=\frac{C-V_{Nor.}S_{Nor.}-V_{Trop.}S_{Trop.}-V_{S}S_{S}-V_{B}S_{B}}{V_{IP}}
\end{equation}
Here, $V_{i}$ denotes the volume of each box, $\gamma$ is the proportion of the cold water path; $K_{i}$ represent wind driven transports between surface boxes, $F_{i}$ are net surface freshwater fluxes. The $S_{0}$ is the reference salinity, set to 35 psu.

To simulate internal climate variability, we incorporate stochastic perturbations in the form of additive white noise:
\begin{equation}
    dx(t)=f(x(t),t)dt+BdW_t
\end{equation}
where $x(t)=(S_{Nor.},S_{Trop.})$ is a vector that represents the salinities in Northern Atlantic and Tropical Atlantic boxes at a time step $t$ and $f$ is the deterministic box model calibrated to the piControl output from HadGEM3. $B$ is a constant matrix of driving noise amplitudes and $W_t$ is a vector of a standard independent Wiener process (each component of $W_t$ is a Brownian motion with zero mean).
Parameter values used in the simulations are summarized in Supplementary Table S1, based on those in Alkhayuon et al. \cite{alkhayuon2019basin}, updated according to HadGEM3 \cite{chapman2024tipping, chapman2024quantifying}.

\subsection{Data and preprocessing}\label{subsec3}
To comprehensively investigate the tipping behavior of the AMOC, we conducted extensive simulations across wide ranges of freshwater forcing magnitudes and rates, each subjected to stochastic perturbations in the form of additive white noise. As illustrated in Fig. 1e, the freshwater forcing magnitudes ranged from 0.34 to 0.46 Sv, with higher resolution (0.005 Sv intervals) between 0.36 and 0.43 Sv to better capture critical transitions. In each scenario, the forcing increased linearly over a designated ramp-up period ranging from 50 to 1500 years in 50-year intervals, then remained fixed at its target value for an additional 2,000 years to allow the system to equilibrate. Each scenario was simulated 20,000 times with distinct noise realizations to ensure statistical robustness. In our algorithm to count the tipping proportion, the tipping point is defined as a persistent drop in AMOC strength below 0 Sv, which clearly separates irreversible collapse from transient weakening. This approach improves upon earlier tipping criteria (e.g., AMOC strength below 5 Sv \cite{chapman2024tipping}), which we found often included self-recovering trajectories that confound the subsequent analysis. 

We first train a CNN-based model using data from a representative scenario, in which forcing increased from 0 to 0.42 Sv over 800-year. This scenario yields a balanced dataset of 100,000 time series: 50,000 that exhibited AMOC collapse and 50,000 that did not. To identify dynamic and higher-order temporal precursors to the tipping of the AMOC, all time series are detrended and temporally aligned relative to the tipping point: For each time series exhibiting a tipping event, we truncate the series at the tipping point and retain the segment preceding it. This enables a focused analysis of the system’s behavior leading up to the critical transition. For each time series without a tipping event, we randomly assign a pseudo-tipping point by sampling from the distribution of actual tipping times in the dataset. The time series is then truncated at that assigned point, and the preceding segment is retained. All truncated segments -- whether from tipping or non-tipping cases -- are subsequently aligned at time 0 to explore common patterns preceding the occurrence of AMOC tipping. This dataset is subsequently divided into train, validation, and test using an 8:1:1 ratio to enable robust model training and evaluation. To highlight intrinsic fluctuations and suppress long-term trends, each time series was detrended by subtracting the ensemble mean of the non-tipping trajectories over the entire time span. This approach allows us to isolate early warning signals relative to a stable baseline that is unaffected by collapse dynamics.

\subsection{Computation and evaluation of CSD indicators}\label{subsec4}
Autocorrelation and variance are computed for all time series following a same procedure.  Each time series undergoes nonlinear detrending using ordinary least squares regression to remove long-term trends. Subsequently, the CSD indicators lag-1 autocorrelation and variance, are computed within a sliding window of 200 time steps, where both indicators are calculated using the standard way \cite{boers2022theoretical, scheffer2009early}. We experiment with different sizes for the sliding windows, ranging from 50 to 400 time steps. The resulting CSD indicators do not show significant divergence, and the insights for the inferences remain consistent.

To quantitatively compare the predictive performance of the CSD indicators with our CNN-based classifier, we employ the area under the receiver operating characteristic (ROC) curve (AUC) as the evaluation metric \cite{FAWCETT2006861}. The ROC curve captures the trade-off between true positive and false positive rates across varying discrimination thresholds. For the CSD indicators, the Kendall $\tau$ coefficient is used to summarize the temporal trend of the indicator and served as the discrimination threshold. For the AMOC strength indicator, its raw value is used directly as the threshold variable. For the CNN-based classifier, the predicted tipping probabilities are used directly as the threshold variable. This unified evaluation framework enables a fair and consistent comparison between classical early-warning indicators and the CNN-based approach.

\subsection{Configurations for deep learning}\label{subsec5}
We design and train a CNN model to predict the probability of AMOC tipping in real time. As illustrated in Supplementary Fig. S6, the architecture comprises two convolutional blocks. The first block contains two convolutional layers with a kernel size (3,1) and 32 output channels, each followed by a ReLU activation and a (3,1) max-pooling layer. The second block includes two convolutional layers with a kernel size (3,3) and 64 channels, also followed by ReLU activations and a (3,1) max-pooling layer. To reduce overfitting, a dropout layer with a rate of 0.4 is applied. The output feature maps are flattened and passed through two fully connected layers, with the final layer producing tipping probabilities via a softmax activation. 

The model is implemented in PyTorch 2.5.0 and trained for up to 300 epochs with a batch size of 64, using the cross-entropy loss function and stochastic gradient descent (SGD) optimizer. The initial learning rate was set to 0.01 and decayed to 0.001 via a cosine annealing schedule. A weight decay of 1e-4 was applied for regularization. Learning rate adjustments were performed with a patience of 5 epochs, and early stopping is triggered if validation performance did not improve for 50 consecutive epochs. These training configurations are applied consistently across all models in this study.

Model performance is evaluated using classification accuracy, defined as the proportion of correctly classified tipping and non-tipping cases. A prediction is considered correct if the model assigned a tipping probability greater than 0.5 to a tipping case, or less than 0.5 to a non-tipping case.

\subsection*{Data availability}
The data involved in the study will be deposited in the Zenodo public repository after the manuscript is accepted.

\subsection*{Code availability}
All code involved in the study will be deposited in the Zenodo and Github public repository after the manuscript is accepted.

\bibliography{sn-bibliography}

\begin{thebibliography}{10}
\expandafter\ifx\csname url\endcsname\relax
  \def\url#1{\burl{#1}}\fi
\expandafter\ifx\csname urlprefix\endcsname\relax\def\urlprefix{URL }\fi
\providecommand{\bibinfo}[2]{#2}
\providecommand{\eprint}[2][]{\url{#2}}
\providecommand{\doi}[1]{\url{https://doi.org/#1}}
\bibcommenthead

\bibitem{dakos2008slowing}
\bibinfo{author}{Dakos, V.} \emph{et~al.}
\newblock \bibinfo{title}{Slowing down as an early warning signal for abrupt
  climate change}.
\newblock \emph{\bibinfo{journal}{Proceedings of the National Academy of
  Sciences}} \textbf{\bibinfo{volume}{105}}, \bibinfo{pages}{14308--14312}
  (\bibinfo{year}{2008}).

\bibitem{armstrong2022exceeding}
\bibinfo{author}{Armstrong~McKay, D.~I.} \emph{et~al.}
\newblock \bibinfo{title}{Exceeding 1.5 c global warming could trigger multiple
  climate tipping points}.
\newblock \emph{\bibinfo{journal}{Science}} \textbf{\bibinfo{volume}{377}},
  \bibinfo{pages}{eabn7950} (\bibinfo{year}{2022}).

\bibitem{boers2022theoretical}
\bibinfo{author}{Boers, N.}, \bibinfo{author}{Ghil, M.} \&
  \bibinfo{author}{Stocker, T.~F.}
\newblock \bibinfo{title}{Theoretical and paleoclimatic evidence for abrupt
  transitions in the earth system}.
\newblock \emph{\bibinfo{journal}{Environmental Research Letters}}
  \textbf{\bibinfo{volume}{17}}, \bibinfo{pages}{093006}
  (\bibinfo{year}{2022}).

\bibitem{flores2024critical}
\bibinfo{author}{Flores, B.~M.} \emph{et~al.}
\newblock \bibinfo{title}{Critical transitions in the amazon forest system}.
\newblock \emph{\bibinfo{journal}{Nature}} \textbf{\bibinfo{volume}{626}},
  \bibinfo{pages}{555--564} (\bibinfo{year}{2024}).

\bibitem{boulton2022pronounced}
\bibinfo{author}{Boulton, C.~A.}, \bibinfo{author}{Lenton, T.~M.} \&
  \bibinfo{author}{Boers, N.}
\newblock \bibinfo{title}{Pronounced loss of amazon rainforest resilience since
  the early 2000s}.
\newblock \emph{\bibinfo{journal}{Nature Climate Change}}
  \textbf{\bibinfo{volume}{12}}, \bibinfo{pages}{271--278}
  (\bibinfo{year}{2022}).

\bibitem{garbe2020hysteresis}
\bibinfo{author}{Garbe, J.}, \bibinfo{author}{Albrecht, T.},
  \bibinfo{author}{Levermann, A.}, \bibinfo{author}{Donges, J.~F.} \&
  \bibinfo{author}{Winkelmann, R.}
\newblock \bibinfo{title}{The hysteresis of the antarctic ice sheet}.
\newblock \emph{\bibinfo{journal}{Nature}} \textbf{\bibinfo{volume}{585}},
  \bibinfo{pages}{538--544} (\bibinfo{year}{2020}).

\bibitem{boers2021observation}
\bibinfo{author}{Boers, N.}
\newblock \bibinfo{title}{Observation-based early-warning signals for a
  collapse of the atlantic meridional overturning circulation}.
\newblock \emph{\bibinfo{journal}{Nature Climate Change}}
  \textbf{\bibinfo{volume}{11}}, \bibinfo{pages}{680--688}
  (\bibinfo{year}{2021}).

\bibitem{van2024physics}
\bibinfo{author}{Van~Westen, R.~M.}, \bibinfo{author}{Kliphuis, M.} \&
  \bibinfo{author}{Dijkstra, H.~A.}
\newblock \bibinfo{title}{Physics-based early warning signal shows that amoc is
  on tipping course}.
\newblock \emph{\bibinfo{journal}{Science advances}}
  \textbf{\bibinfo{volume}{10}}, \bibinfo{pages}{eadk1189}
  (\bibinfo{year}{2024}).

\bibitem{ashwin2012tipping}
\bibinfo{author}{Ashwin, P.}, \bibinfo{author}{Wieczorek, S.},
  \bibinfo{author}{Vitolo, R.} \& \bibinfo{author}{Cox, P.}
\newblock \bibinfo{title}{Tipping points in open systems: bifurcation,
  noise-induced and rate-dependent examples in the climate system}.
\newblock \emph{\bibinfo{journal}{Philosophical Transactions of the Royal
  Society A: Mathematical, Physical and Engineering Sciences}}
  \textbf{\bibinfo{volume}{370}}, \bibinfo{pages}{1166--1184}
  (\bibinfo{year}{2012}).

\bibitem{brunetti2023attractors}
\bibinfo{author}{Brunetti, M.} \& \bibinfo{author}{Ragon, C.}
\newblock \bibinfo{title}{Attractors and bifurcation diagrams in complex
  climate models}.
\newblock \emph{\bibinfo{journal}{Physical Review E}}
  \textbf{\bibinfo{volume}{107}}, \bibinfo{pages}{054214}
  (\bibinfo{year}{2023}).

\bibitem{chapman2024tipping}
\bibinfo{author}{Chapman, R.}, \bibinfo{author}{Sinet, S.} \&
  \bibinfo{author}{Ritchie, P.~D.}
\newblock \bibinfo{title}{Tipping mechanisms in a conceptual model of the
  atlantic meridional overturning circulation}.
\newblock \emph{\bibinfo{journal}{Weather}} \textbf{\bibinfo{volume}{79}},
  \bibinfo{pages}{316--323} (\bibinfo{year}{2024}).

\bibitem{scheffer2009early}
\bibinfo{author}{Scheffer, M.} \emph{et~al.}
\newblock \bibinfo{title}{Early-warning signals for critical transitions}.
\newblock \emph{\bibinfo{journal}{Nature}} \textbf{\bibinfo{volume}{461}},
  \bibinfo{pages}{53--59} (\bibinfo{year}{2009}).

\bibitem{Kleinen2003}
\bibinfo{author}{Kleinen, T.}, \bibinfo{author}{Held, H.} \&
  \bibinfo{author}{Petschel-Held, G.}
\newblock \bibinfo{title}{The potential role of spectral properties in
  detecting thresholds in the earth system: application to the thermohaline
  circulation}.
\newblock \emph{\bibinfo{journal}{Ocean Dynamics}}
  \textbf{\bibinfo{volume}{53}}, \bibinfo{pages}{53–63}
  (\bibinfo{year}{2003}).

\bibitem{held2004detection}
\bibinfo{author}{Held, H.} \& \bibinfo{author}{Kleinen, T.}
\newblock \bibinfo{title}{Detection of climate system bifurcations by
  degenerate fingerprinting}.
\newblock \emph{\bibinfo{journal}{Geophysical Research Letters}}
  \textbf{\bibinfo{volume}{31}} (\bibinfo{year}{2004}).

\bibitem{lohmann2021risk}
\bibinfo{author}{Lohmann, J.} \& \bibinfo{author}{Ditlevsen, P.~D.}
\newblock \bibinfo{title}{Risk of tipping the overturning circulation due to
  increasing rates of ice melt}.
\newblock \emph{\bibinfo{journal}{Proceedings of the National Academy of
  Sciences}} \textbf{\bibinfo{volume}{118}}, \bibinfo{pages}{e2017989118}
  (\bibinfo{year}{2021}).

\bibitem{alharthi2023rate}
\bibinfo{author}{Alharthi, T.}
\newblock \emph{\bibinfo{title}{Rate-dependent tipping points for forced
  dynamical systems}}.
\newblock Ph.D. thesis, \bibinfo{school}{University of Exeter}
  (\bibinfo{year}{2023}).

\bibitem{huang2024deep}
\bibinfo{author}{Huang, Y.}, \bibinfo{author}{Bathiany, S.},
  \bibinfo{author}{Ashwin, P.} \& \bibinfo{author}{Boers, N.}
\newblock \bibinfo{title}{Deep learning for predicting rate-induced tipping}.
\newblock \emph{\bibinfo{journal}{Nature Machine Intelligence}}
  \bibinfo{pages}{1--10} (\bibinfo{year}{2024}).

\bibitem{hasselmann1976stochastic}
\bibinfo{author}{Hasselmann, K.}
\newblock \bibinfo{title}{Stochastic climate models part i. theory}.
\newblock \emph{\bibinfo{journal}{tellus}} \textbf{\bibinfo{volume}{28}},
  \bibinfo{pages}{473--485} (\bibinfo{year}{1976}).

\bibitem{timmermann2000noise}
\bibinfo{author}{Timmermann, A.} \& \bibinfo{author}{Lohmann, G.}
\newblock \bibinfo{title}{Noise-induced transitions in a simplified model of
  the thermohaline circulation}.
\newblock \emph{\bibinfo{journal}{Journal of Physical Oceanography}}
  \textbf{\bibinfo{volume}{30}}, \bibinfo{pages}{1891--1900}
  (\bibinfo{year}{2000}).

\bibitem{stocker1997influence}
\bibinfo{author}{Stocker, T.~F.} \& \bibinfo{author}{Schmittner, A.}
\newblock \bibinfo{title}{Influence of co2 emission rates on the stability of
  the thermohaline circulation}.
\newblock \emph{\bibinfo{journal}{Nature}} \textbf{\bibinfo{volume}{388}},
  \bibinfo{pages}{862--865} (\bibinfo{year}{1997}).

\bibitem{boers2018early}
\bibinfo{author}{Boers, N.}
\newblock \bibinfo{title}{Early-warning signals for dansgaard-oeschger events
  in a high-resolution ice core record}.
\newblock \emph{\bibinfo{journal}{Nature communications}}
  \textbf{\bibinfo{volume}{9}}, \bibinfo{pages}{2556} (\bibinfo{year}{2018}).

\bibitem{cini2024simulating}
\bibinfo{author}{Cini, M.}, \bibinfo{author}{Zappa, G.},
  \bibinfo{author}{Ragone, F.} \& \bibinfo{author}{Corti, S.}
\newblock \bibinfo{title}{Simulating amoc tipping driven by internal climate
  variability with a rare event algorithm}.
\newblock \emph{\bibinfo{journal}{npj Climate and Atmospheric Science}}
  \textbf{\bibinfo{volume}{7}}, \bibinfo{pages}{31} (\bibinfo{year}{2024}).

\bibitem{lenton2019climate}
\bibinfo{author}{Lenton, T.~M.} \emph{et~al.}
\newblock \bibinfo{title}{Climate tipping points—too risky to bet against}.
\newblock \emph{\bibinfo{journal}{Nature}} \textbf{\bibinfo{volume}{575}},
  \bibinfo{pages}{592--595} (\bibinfo{year}{2019}).

\bibitem{ritchie2021overshooting}
\bibinfo{author}{Ritchie, P.~D.}, \bibinfo{author}{Clarke, J.~J.},
  \bibinfo{author}{Cox, P.~M.} \& \bibinfo{author}{Huntingford, C.}
\newblock \bibinfo{title}{Overshooting tipping point thresholds in a changing
  climate}.
\newblock \emph{\bibinfo{journal}{Nature}} \textbf{\bibinfo{volume}{592}},
  \bibinfo{pages}{517--523} (\bibinfo{year}{2021}).

\bibitem{wunderling2023global}
\bibinfo{author}{Wunderling, N.} \emph{et~al.}
\newblock \bibinfo{title}{Global warming overshoots increase risks of climate
  tipping cascades in a network model}.
\newblock \emph{\bibinfo{journal}{Nature Climate Change}}
  \textbf{\bibinfo{volume}{13}}, \bibinfo{pages}{75--82}
  (\bibinfo{year}{2023}).

\bibitem{romanou2023stochastic}
\bibinfo{author}{Romanou, A.} \emph{et~al.}
\newblock \bibinfo{title}{Stochastic bifurcation of the north atlantic
  circulation under a mid-range future climate scenario with the nasa-giss
  modele}.
\newblock \emph{\bibinfo{journal}{Journal of Climate}} \bibinfo{pages}{1--49}
  (\bibinfo{year}{2023}).

\bibitem{klose2024rate}
\bibinfo{author}{Klose, A.~K.}, \bibinfo{author}{Donges, J.~F.},
  \bibinfo{author}{Feudel, U.} \& \bibinfo{author}{Winkelmann, R.}
\newblock \bibinfo{title}{Rate-induced tipping cascades arising from
  interactions between the greenland ice sheet and the atlantic meridional
  overturning circulation}.
\newblock \emph{\bibinfo{journal}{Earth System Dynamics}}
  \textbf{\bibinfo{volume}{15}}, \bibinfo{pages}{635--652}
  (\bibinfo{year}{2024}).

\bibitem{sinet2024amoc}
\bibinfo{author}{Sinet, S.}, \bibinfo{author}{Ashwin, P.},
  \bibinfo{author}{von~der Heydt, A.~S.} \& \bibinfo{author}{Dijkstra, H.~A.}
\newblock \bibinfo{title}{Amoc stability amid tipping ice sheets: the crucial
  role of rate and noise}.
\newblock \emph{\bibinfo{journal}{Earth System Dynamics}}
  \textbf{\bibinfo{volume}{15}}, \bibinfo{pages}{859--873}
  (\bibinfo{year}{2024}).

\bibitem{baker2025continued}
\bibinfo{author}{Baker, J.} \emph{et~al.}
\newblock \bibinfo{title}{Continued atlantic overturning circulation even under
  climate extremes}.
\newblock \emph{\bibinfo{journal}{Nature}} \textbf{\bibinfo{volume}{638}},
  \bibinfo{pages}{987--994} (\bibinfo{year}{2025}).

\bibitem{ben2024uncertainties}
\bibinfo{author}{Ben-Yami, M.}, \bibinfo{author}{Morr, A.},
  \bibinfo{author}{Bathiany, S.} \& \bibinfo{author}{Boers, N.}
\newblock \bibinfo{title}{Uncertainties too large to predict tipping times of
  major earth system components from historical data}.
\newblock \emph{\bibinfo{journal}{Science Advances}}
  \textbf{\bibinfo{volume}{10}}, \bibinfo{pages}{eadl4841}
  (\bibinfo{year}{2024}).

\bibitem{rietkerk2025ambiguity}
\bibinfo{author}{Rietkerk, M.}, \bibinfo{author}{Skiba, V.},
  \bibinfo{author}{Weinans, E.}, \bibinfo{author}{H{\'e}bert, R.} \&
  \bibinfo{author}{Laepple, T.}
\newblock \bibinfo{title}{Ambiguity of early warning signals for climate
  tipping points}.
\newblock \emph{\bibinfo{journal}{Nature Climate Change}}
  \bibinfo{pages}{1--10} (\bibinfo{year}{2025}).

\bibitem{zhang2017abrupt}
\bibinfo{author}{Zhang, X.}, \bibinfo{author}{Knorr, G.},
  \bibinfo{author}{Lohmann, G.} \& \bibinfo{author}{Barker, S.}
\newblock \bibinfo{title}{Abrupt north atlantic circulation changes in response
  to gradual co2 forcing in a glacial climate state}.
\newblock \emph{\bibinfo{journal}{Nature Geoscience}}
  \textbf{\bibinfo{volume}{10}}, \bibinfo{pages}{518--523}
  (\bibinfo{year}{2017}).

\bibitem{ben2023uncertainties}
\bibinfo{author}{Ben-Yami, M.}, \bibinfo{author}{Skiba, V.},
  \bibinfo{author}{Bathiany, S.} \& \bibinfo{author}{Boers, N.}
\newblock \bibinfo{title}{Uncertainties in critical slowing down indicators of
  observation-based fingerprints of the atlantic overturning circulation}.
\newblock \emph{\bibinfo{journal}{Nature Communications}}
  \textbf{\bibinfo{volume}{14}}, \bibinfo{pages}{8344} (\bibinfo{year}{2023}).

\bibitem{ditlevsen2023warning}
\bibinfo{author}{Ditlevsen, P.} \& \bibinfo{author}{Ditlevsen, S.}
\newblock \bibinfo{title}{Warning of a forthcoming collapse of the atlantic
  meridional overturning circulation}.
\newblock \emph{\bibinfo{journal}{Nature Communications}}
  \textbf{\bibinfo{volume}{14}}, \bibinfo{pages}{1--12} (\bibinfo{year}{2023}).

\bibitem{dijkstra2024role}
\bibinfo{author}{Dijkstra, H.~A.}
\newblock \bibinfo{title}{The role of conceptual models in climate research}.
\newblock \emph{\bibinfo{journal}{Physica D: Nonlinear Phenomena}}
  \textbf{\bibinfo{volume}{457}}, \bibinfo{pages}{133984}
  (\bibinfo{year}{2024}).

\bibitem{lohmann2024multistability}
\bibinfo{author}{Lohmann, J.}, \bibinfo{author}{Dijkstra, H.~A.},
  \bibinfo{author}{Jochum, M.}, \bibinfo{author}{Lucarini, V.} \&
  \bibinfo{author}{Ditlevsen, P.~D.}
\newblock \bibinfo{title}{Multistability and intermediate tipping of the
  atlantic ocean circulation}.
\newblock \emph{\bibinfo{journal}{Science advances}}
  \textbf{\bibinfo{volume}{10}}, \bibinfo{pages}{eadi4253}
  (\bibinfo{year}{2024}).

\bibitem{gu2024wide}
\bibinfo{author}{Gu, Q.} \emph{et~al.}
\newblock \bibinfo{title}{Wide range of possible trajectories of north atlantic
  climate in a warming world}.
\newblock \emph{\bibinfo{journal}{Nature communications}}
  \textbf{\bibinfo{volume}{15}}, \bibinfo{pages}{4221} (\bibinfo{year}{2024}).

\bibitem{curtis2024collapse}
\bibinfo{author}{Curtis, P.~E.} \& \bibinfo{author}{Fedorov, A.~V.}
\newblock \bibinfo{title}{Collapse and slow recovery of the atlantic meridional
  overturning circulation (amoc) under abrupt greenhouse gas forcing}.
\newblock \emph{\bibinfo{journal}{Climate Dynamics}}
  \textbf{\bibinfo{volume}{62}}, \bibinfo{pages}{5949--5970}
  (\bibinfo{year}{2024}).

\bibitem{frierson2013contribution}
\bibinfo{author}{Frierson, D.~M.} \emph{et~al.}
\newblock \bibinfo{title}{Contribution of ocean overturning circulation to
  tropical rainfall peak in the northern hemisphere}.
\newblock \emph{\bibinfo{journal}{Nature Geoscience}}
  \textbf{\bibinfo{volume}{6}}, \bibinfo{pages}{940--944}
  (\bibinfo{year}{2013}).

\bibitem{wood2019observable}
\bibinfo{author}{Wood, R.~A.}, \bibinfo{author}{Rodr{\'\i}guez, J.~M.},
  \bibinfo{author}{Smith, R.~S.}, \bibinfo{author}{Jackson, L.~C.} \&
  \bibinfo{author}{Hawkins, E.}
\newblock \bibinfo{title}{Observable, low-order dynamical controls on
  thresholds of the atlantic meridional overturning circulation}.
\newblock \emph{\bibinfo{journal}{Climate Dynamics}}
  \textbf{\bibinfo{volume}{53}}, \bibinfo{pages}{6815--6834}
  (\bibinfo{year}{2019}).

\bibitem{alkhayuon2019basin}
\bibinfo{author}{Alkhayuon, H.}, \bibinfo{author}{Ashwin, P.},
  \bibinfo{author}{Jackson, L.~C.}, \bibinfo{author}{Quinn, C.} \&
  \bibinfo{author}{Wood, R.~A.}
\newblock \bibinfo{title}{Basin bifurcations, oscillatory instability and
  rate-induced thresholds for atlantic meridional overturning circulation in a
  global oceanic box model}.
\newblock \emph{\bibinfo{journal}{Proceedings of the Royal Society A}}
  \textbf{\bibinfo{volume}{475}}, \bibinfo{pages}{20190051}
  (\bibinfo{year}{2019}).

\bibitem{chapman2024quantifying}
\bibinfo{author}{Chapman, R.}, \bibinfo{author}{Ashwin, P.},
  \bibinfo{author}{Baker, J.} \& \bibinfo{author}{Wood, R.}
\newblock \bibinfo{title}{Quantifying risk of a noise-induced amoc collapse
  from northern and tropical atlantic ocean variability}.
\newblock \emph{\bibinfo{journal}{Environmental Research Communications}}
  \textbf{\bibinfo{volume}{6}}, \bibinfo{pages}{111003} (\bibinfo{year}{2024}).

\bibitem{panahi2024machine}
\bibinfo{author}{Panahi, S.} \emph{et~al.}
\newblock \bibinfo{title}{Machine learning prediction of tipping in complex
  dynamical systems}.
\newblock \emph{\bibinfo{journal}{Physical Review Research}}
  \textbf{\bibinfo{volume}{6}}, \bibinfo{pages}{043194} (\bibinfo{year}{2024}).

\bibitem{pals2024targeted}
\bibinfo{author}{Pals, D.}, \bibinfo{author}{Bathiany, S.},
  \bibinfo{author}{Wood, R.} \& \bibinfo{author}{Boers, N.}
\newblock \bibinfo{title}{Targeted calibration to adjust stability biases in
  non-differentiable complex system models}.
\newblock \emph{\bibinfo{journal}{arXiv preprint arXiv:2409.04063}}
  (\bibinfo{year}{2024}).

\bibitem{lenton2011early}
\bibinfo{author}{Lenton, T.~M.}
\newblock \bibinfo{title}{Early warning of climate tipping points}.
\newblock \emph{\bibinfo{journal}{Nature climate change}}
  \textbf{\bibinfo{volume}{1}}, \bibinfo{pages}{201--209}
  (\bibinfo{year}{2011}).

\bibitem{bury2021deep}
\bibinfo{author}{Bury, T.~M.} \emph{et~al.}
\newblock \bibinfo{title}{Deep learning for early warning signals of tipping
  points}.
\newblock \emph{\bibinfo{journal}{Proceedings of the National Academy of
  Sciences}} \textbf{\bibinfo{volume}{118}}, \bibinfo{pages}{e2106140118}
  (\bibinfo{year}{2021}).

\bibitem{bury2023predicting}
\bibinfo{author}{Bury, T.~M.} \emph{et~al.}
\newblock \bibinfo{title}{Predicting discrete-time bifurcations with deep
  learning}.
\newblock \emph{\bibinfo{journal}{Nature Communications}}
  \textbf{\bibinfo{volume}{14}}, \bibinfo{pages}{6331} (\bibinfo{year}{2023}).

\bibitem{FAWCETT2006861}
\bibinfo{author}{Fawcett, T.}
\newblock \bibinfo{title}{An introduction to roc analysis}.
\newblock \emph{\bibinfo{journal}{Pattern Recognition Letters}}
  \textbf{\bibinfo{volume}{27}}, \bibinfo{pages}{861--874}
  (\bibinfo{year}{2006}).

\bibitem{scholkopf1997kernel}
\bibinfo{author}{Sch{\"o}lkopf, B.}, \bibinfo{author}{Smola, A.} \&
  \bibinfo{author}{M{\"u}ller, K.-R.}
\newblock \bibinfo{title}{Kernel principal component analysis}.
\newblock \emph{\bibinfo{journal}{International conference on artificial neural
  networks}} \bibinfo{pages}{583--588} (\bibinfo{year}{1997}).

\bibitem{he2013detecting}
\bibinfo{author}{He, W.} \emph{et~al.}
\newblock \bibinfo{title}{Detecting abrupt change on the basis of skewness:
  numerical tests and applications.}
\newblock \emph{\bibinfo{journal}{International Journal of Climatology}}
  \textbf{\bibinfo{volume}{33}} (\bibinfo{year}{2013}).

\bibitem{bach2015pixel}
\bibinfo{author}{Bach, S.} \emph{et~al.}
\newblock \bibinfo{title}{On pixel-wise explanations for non-linear classifier
  decisions by layer-wise relevance propagation}.
\newblock \emph{\bibinfo{journal}{PloS one}} \textbf{\bibinfo{volume}{10}},
  \bibinfo{pages}{e0130140} (\bibinfo{year}{2015}).

\bibitem{montavon2019layer}
\bibinfo{author}{Montavon, G.}, \bibinfo{author}{Binder, A.},
  \bibinfo{author}{Lapuschkin, S.}, \bibinfo{author}{Samek, W.} \&
  \bibinfo{author}{M{\"u}ller, K.-R.}
\newblock \bibinfo{title}{Layer-wise relevance propagation: an overview}.
\newblock \emph{\bibinfo{journal}{Explainable AI: interpreting, explaining and
  visualizing deep learning}} \bibinfo{pages}{193--209} (\bibinfo{year}{2019}).

\bibitem{bochow2023overshooting}
\bibinfo{author}{Bochow, N.} \emph{et~al.}
\newblock \bibinfo{title}{Overshooting the critical threshold for the greenland
  ice sheet}.
\newblock \emph{\bibinfo{journal}{Nature}} \textbf{\bibinfo{volume}{622}},
  \bibinfo{pages}{528--536} (\bibinfo{year}{2023}).

\bibitem{shin2024fast}
\bibinfo{author}{Shin, Y.} \emph{et~al.}
\newblock \bibinfo{title}{Fast and slow responses of atlantic meridional
  overturning circulation to antarctic meltwater forcing}.
\newblock \emph{\bibinfo{journal}{Geophysical Research Letters}}
  \textbf{\bibinfo{volume}{51}}, \bibinfo{pages}{e2024GL108272}
  (\bibinfo{year}{2024}).

\bibitem{spirtes2016causal}
\bibinfo{author}{Spirtes, P.} \& \bibinfo{author}{Zhang, K.}
\newblock \bibinfo{title}{Causal discovery and inference: concepts and recent
  methodological advances}.
\newblock \emph{\bibinfo{journal}{Applied informatics}}
  \textbf{\bibinfo{volume}{3}}, \bibinfo{pages}{1--28} (\bibinfo{year}{2016}).

\bibitem{RAISSI2019686}
\bibinfo{author}{Raissi, M.}, \bibinfo{author}{Perdikaris, P.} \&
  \bibinfo{author}{Karniadakis, G.}
\newblock \bibinfo{title}{Physics-informed neural networks: A deep learning
  framework for solving forward and inverse problems involving nonlinear
  partial differential equations}.
\newblock \emph{\bibinfo{journal}{Journal of Computational Physics}}
  \textbf{\bibinfo{volume}{378}}, \bibinfo{pages}{686--707}
  (\bibinfo{year}{2019}).

\end{thebibliography}

\subsection*{Acknowledgements}
This is ClimTip contribution \#XX; the ClimTip project has received funding from the European Union's Horizon Europe research and innovation programme under grant agreement No. 101137601: Funded by the European Union. W.Z. acknowledges funding by the China Scholarship Council. Y.H. also acknowledges Alexander von Humboldt Foundation for Humboldt Research Fellowship. N.B. also acknowledges funding by the Volkswagen foundation. Y.S. is supported by the National Research Foundation of Korea (NRF) grant funded by the Korea government (MSIT) (RS-2022-NR070706 ; RS-2024-00334637).

\subsection*{Author contributions}
Y.H., S.B., W.Z., Y.S. and N.B. conceived and designed the research. W.Z. performed the numerical analysis. All authors interpreted and discussed the results. W.Z. performed the visualization. W.Z. and Y.H. wrote the manuscript with input from all authors.

\subsection*{Competing interests}
The authors declare no competing interests.

\begin{figure}[h]
\centering
\includegraphics[trim=120 0 120 0, clip, width=1\textwidth]{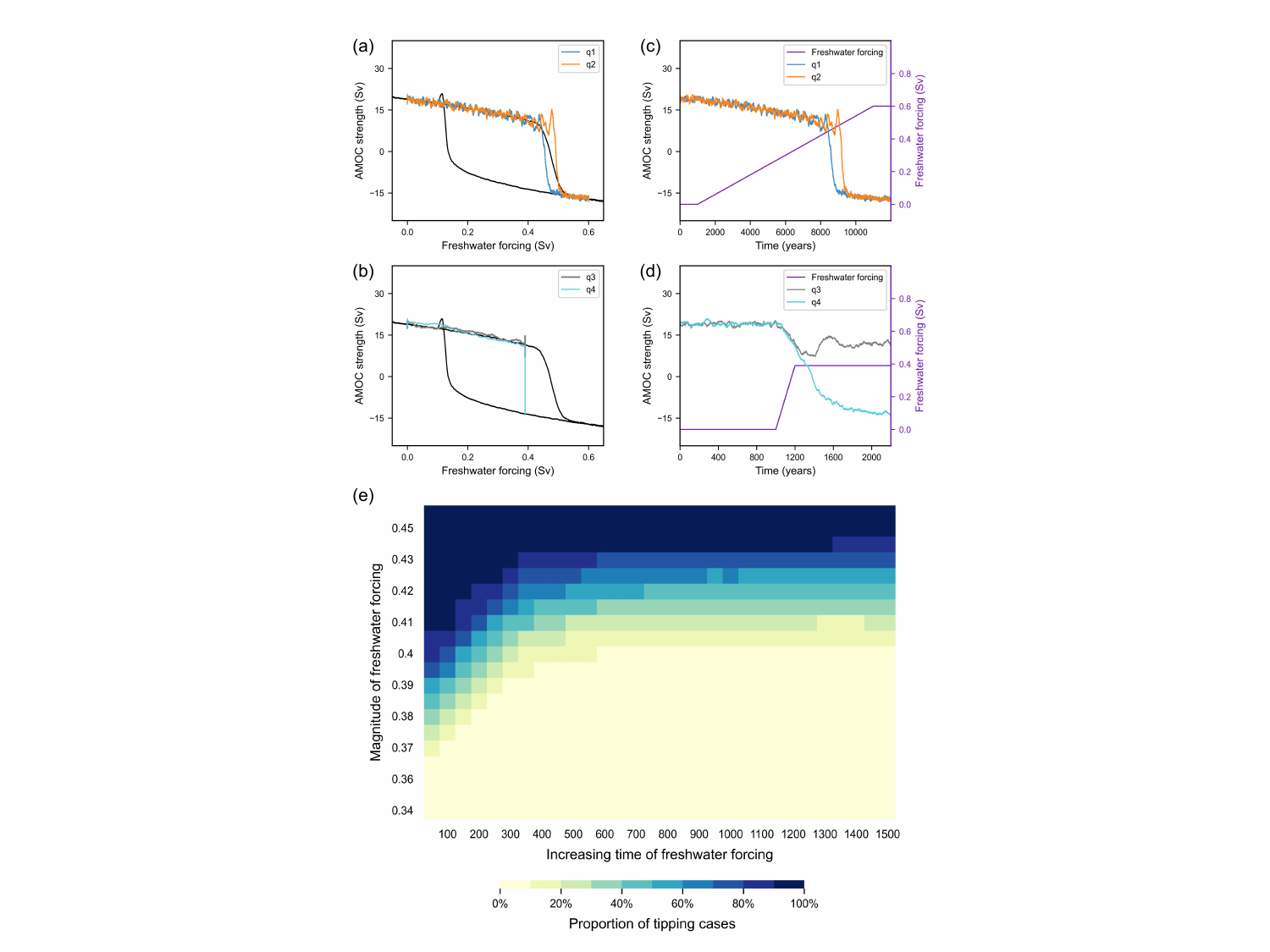}
\caption{Bistability, path-dependence, and collapse likelihood of the AMOC under freshwater forcing, based on simulations of the box model considered here \cite{wood2019observable,chapman2024quantifying}. (a, b) Bifurcation structure of the AMOC in response to freshwater forcing. The black line indicates the bistable regime of the AMOC strength as freshwater forcing linearly increases and then decreases, which is derived from an ensemble average over 50 realizations. Colored trajectories correspond to the transient responses shown in panels (c) and (d).  (c) Two example trajectories (blue and orange) under a linear increase in freshwater forcing to 0.6 Sv over 10,000 years (purple line), both exhibiting collapse, indicative of  bifurcation-induced tipping. (d) Two example trajectories under a faster linear increase in freshwater forcing to 0.39 Sv over 200 years (purple line). The gray trajectory remains in the strong AMOC state, while the cyan trajectory collapses, demonstrating rate-induced tipping. (e) Heat map showing the proportion of AMOC tipping cases across a range of freshwater forcing amplitudes and  times over which the freshwater forcing increases. At each parameter setting, 20,000 stochastic simulations were performed under white noise perturbations. The AMOC tipping is defined as a persistent decline of the AMOC strength below 0 Sv, distinguishing it from temporary or reversible weakening. Color intensity indicates the tipping proportion, highlighting the combined influence of forcing value and rate on AMOC stability.
}\label{fig1}
\end{figure}

\begin{figure}[h]
\centering
\includegraphics[trim=0 0 0 0, clip, width=1\textwidth]{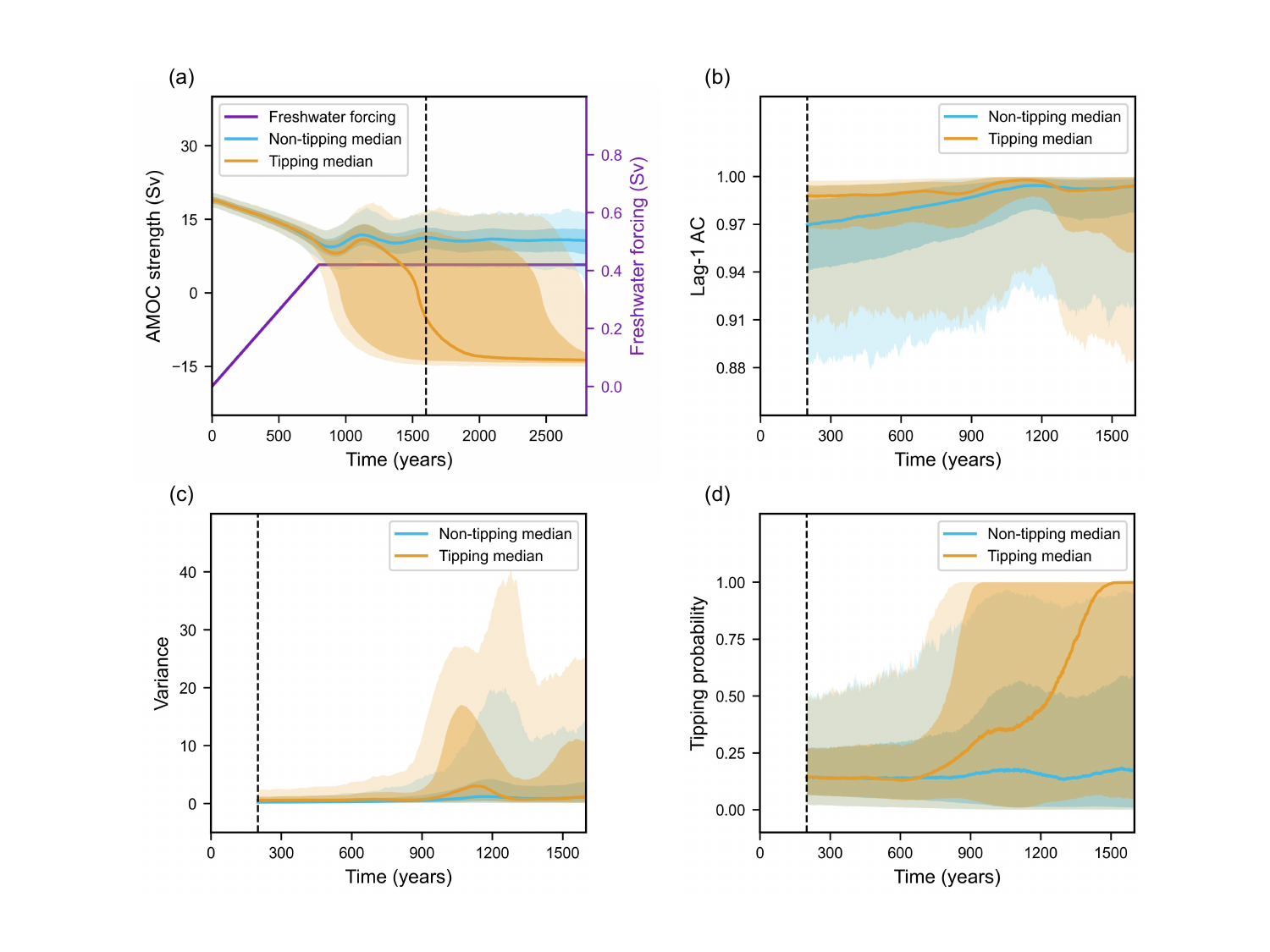}
\caption{Real-time evolution of early warning indicators in the AMOC three-box model. (a) Time series trajectories generated under a linearly increasing freshwater forcing up to 0.42 Sv over 800 years (purple line). The orange and blue lines denote the ensemble means of trajectories that do and do not undergo tipping, respectively. Shaded regions indicate the 75\% (dark) and 99\% (light) confidence intervals. (b) estimated Lag-1 autocorrelation. (c) estimated variance. (d) tipping probabilities generated by the deep learning model. All indicators are computed using a rolling window of 200 time steps.
}\label{fig2}
\end{figure}

\begin{figure}[h]
\centering
\includegraphics[width=1\textwidth]{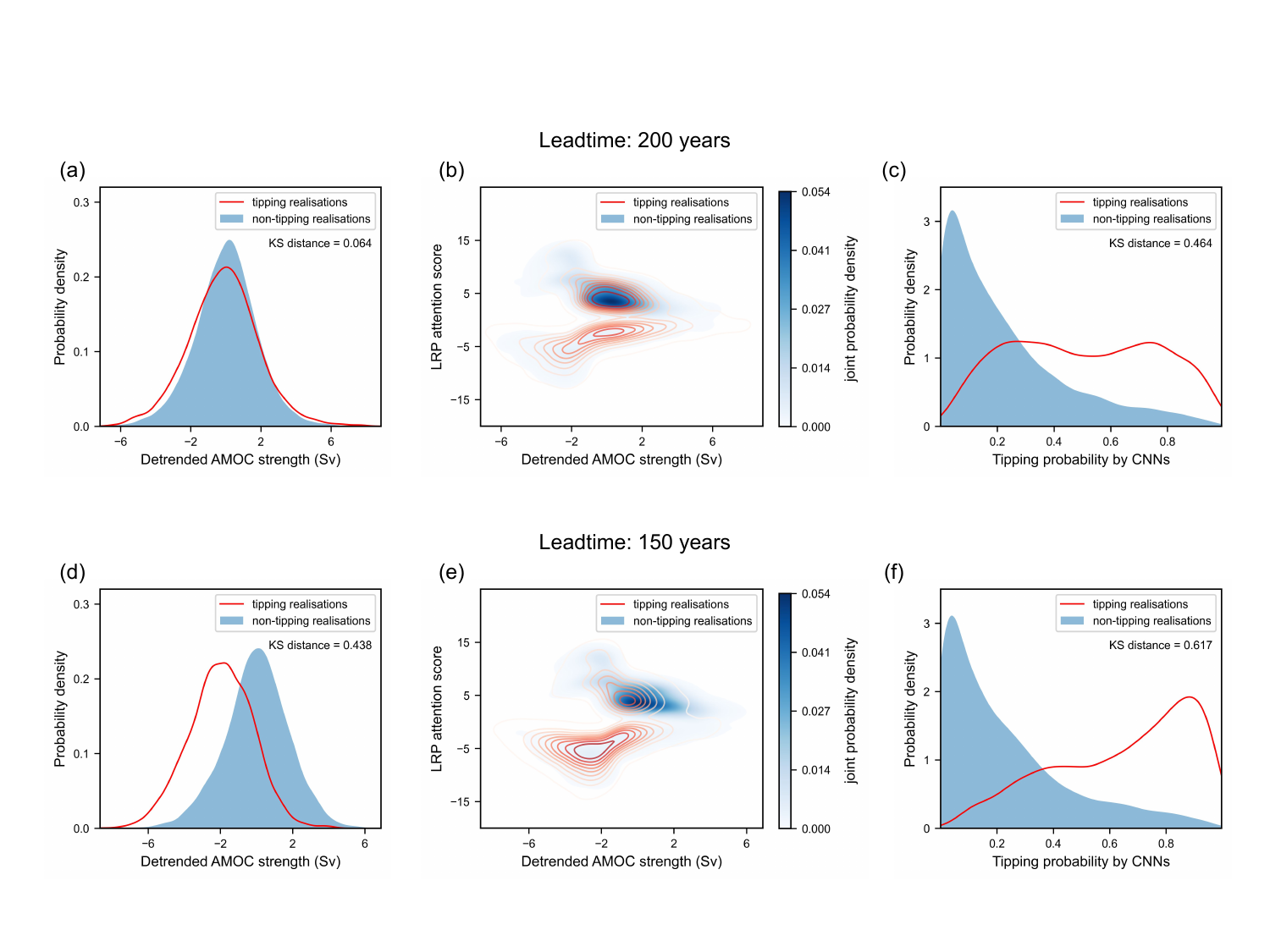}
\caption{Probability distributions of detrended AMOC strength, LRP attention score of CNNs, and tipping probability by CNNs for tipping and non-tipping realizations.
Probability density functions are presented for three indicators: Detrended AMOC strength (a, d), LRP attention score of CNNs (b, e), and the tipping probability by CNNs (c, f). The distributions are shown for both tipping and non-tipping realizations at two forecast leadtimes: 200 years (a-c) and 150 years (d-f). The Kolmogorov–Smirnov (KS) distance, measuring the maximum distance between the cumulative distributions of the two ensembles, is reported for each panel. Higher KS distance indicate better distinguishing ability between tipping and non-tipping scenarios.
}\label{fig3}
\end{figure}

\begin{figure}[h]
\centering
\includegraphics[trim=0 50 0 50, clip, width=1\textwidth]{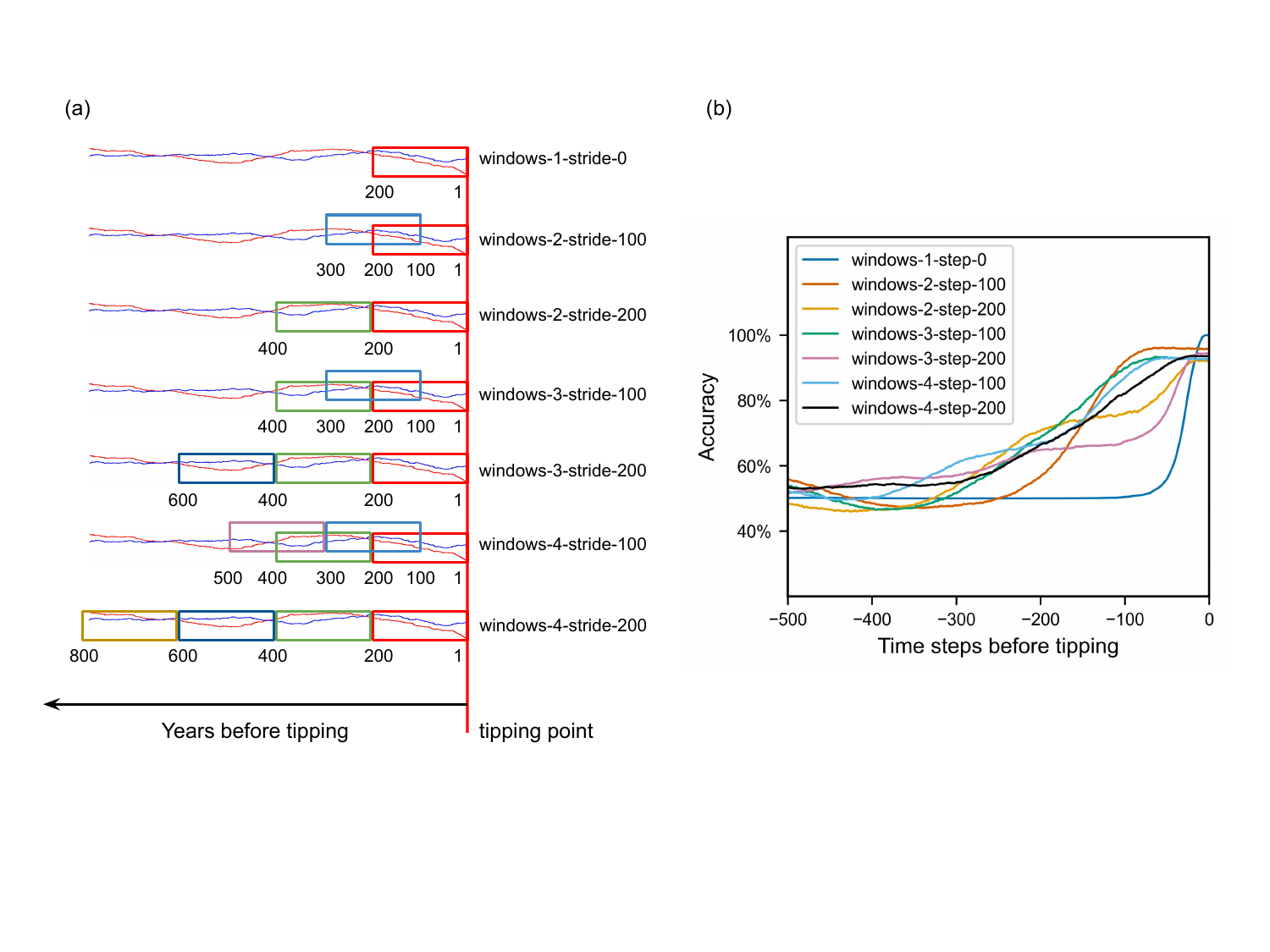}
\caption{(a) Schematic illustration of the sliding-window configurations used to train the CNN classifier. Seven different strategies are presented, each varying in the number of windows and stride lengths. The vertical red line marks the tipping point, and the colored boxes represent 200-year time series segments located at different temporal distances before the tipping point. Configurations include a single window (stride 0) as well as two, three, or four windows with strides of 100 or 200 years. During the training phase, a separate DL model is trained for each configuration using the corresponding time series segments. During inference, the trained model is applied to arbitrary 200-year segments from unseen time series to estimate tipping probabilities at different lead times. (b) Prediction accuracy of the trained models under the seven different training configurations. For each point on the x-axis (years before tipping), accuracy is computed as the proportion of correct predictions, calculated by pooling all predictions from the 200-year test segments that terminate at the corresponding lead time.
}\label{fig4}
\end{figure}

\begin{figure}[h]
\centering
\includegraphics[trim=0 120 0 120, clip, width=1\textwidth]{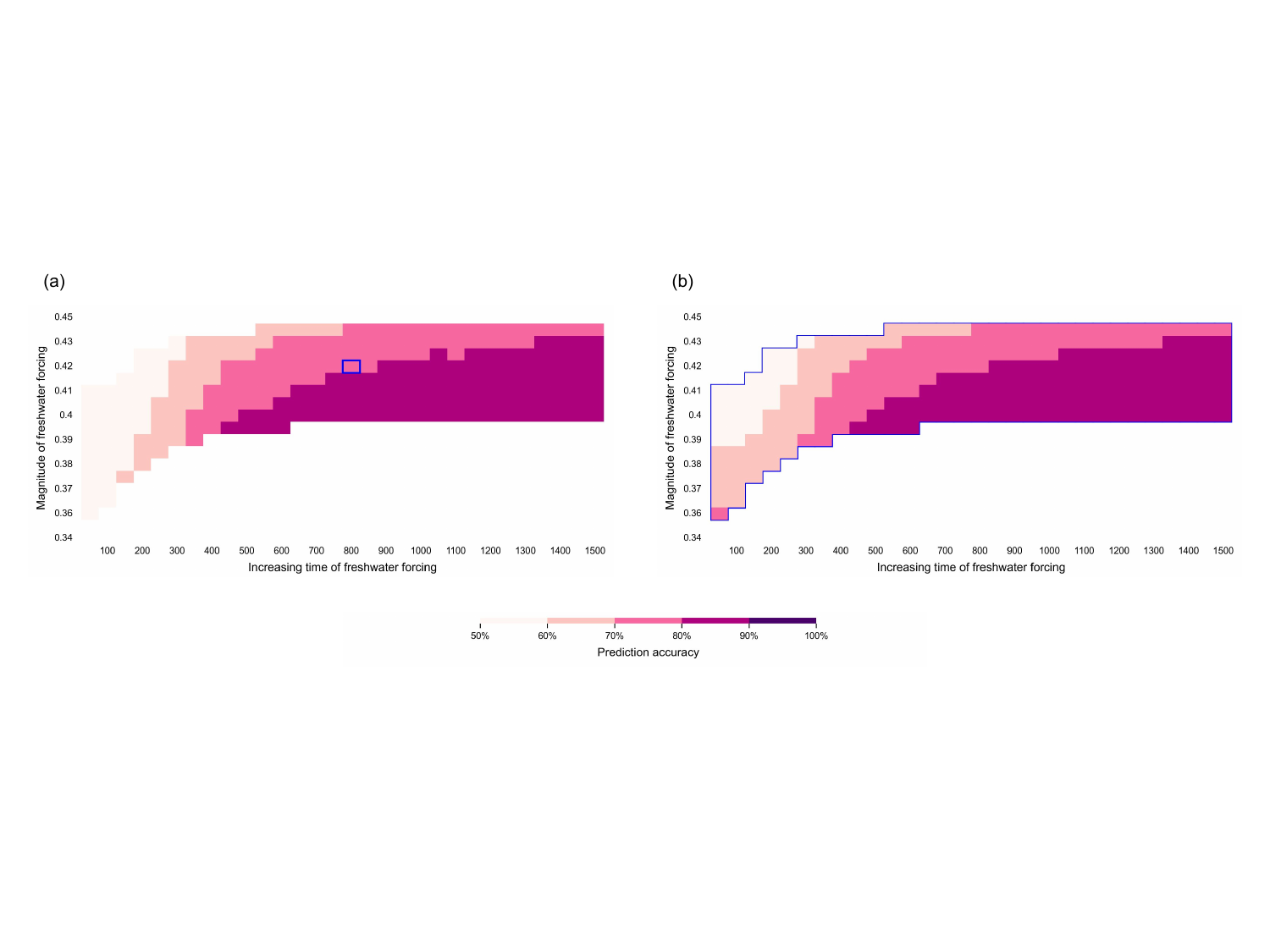}
\caption{Prediction accuracy of the DL model for AMOC tipping across freshwater forcing scenarios. (a, b) Heat maps showing the classification accuracy of the DL model in predicting AMOC tipping using the 150-year trajectory preceding each outcome. Each grid point corresponds to a freshwater forcing scenario, defined by its magnitude and increasing time, consistent with the parameter space in Fig. 1e. Only scenarios with a tipping proportion between 5\% and 95\% were evaluated. (a) DL model was trained on 100,000 trajectories (5,000 tipping and 5,000 non-tipping) generated under a single representative freshwater forcing scenario (linear increase to 0.42 Sv over 800 years). (b) DL model was trained on a scenario-specific, balanced dataset, in which 1,000 tipping and 1,000 non-tipping trajectories were sampled for each evaluated forcing scenario. Color denotes the prediction accuracy for each scenario, revealing the generalization performance of DL models.
}\label{fig5}
\end{figure}

\end{document}